\documentstyle[11pt]{article}  
\newcommand{\be}{\begin{equation}}
\newcommand{\ee}{\end{equation}}

\begin{document}
\title{Non-Hermitian Fermion Mapping for One-Component Plasma}
\author{M. B. Hastings}
\maketitle

\begin{abstract}
The two-dimensional one-component logarithmic Coulomb gas is mapped onto
a non-hermitian fermionic field theory.  At $\beta=2$, the field theory is
free.  Correlation functions are calculated and a perturbation theory is
discussed for extending to other $\beta$.  A phase transition is found at
the mean-field level at large $\beta$.  Some results are extended to spaces of 
constant negative curvature.
   \end{abstract}
\section{Introduction}
The problem of the statistical mechanics of a system of particles in a
two-dimensional plane, all possessing the same sign of charge and
interacting via a repulsive Coulomb interaction, has
been looked at by a number of authors\cite{previous}.  This is
referred to as a one-component plasma since all particles have the same
sign of charge.  The Coulomb interaction
is logarithmic in two-dimensions which makes it possible to introduce
a dimensionless inverse temperature $\beta$ such that the partition function
of an $N$-particle system is equal to 
\be
\int (\prod \limits_{i=1}^{N} dz_i d\overline z_i) \prod \limits_{i<j}^{N} 
|z_i-z_j|^{\beta} \prod \limits_{i=1}^{N} e^{-\beta U(z_i)}
\ee
where $U(r)$ is some background potential and where $z=x+it$ and $\overline z
=x-it$ are coordinates in the complex plane.

It is believed that at some $\beta$ of approximately 140 there is
a first order phase transition in the system\cite{phase}.  Below the phase
transition the system acquires long-range orientational order but does not
acquire long-range positional order.  The system is thus of interest as
a model of crystallization.  
In the last section of the paper, we consider this model on a space of
constant negative curvature.  This may make it possible to look at
effects of frustration since it is not possible to form a hexagonal
crystal lattice without defects on a negative curvature space\cite{neg}.

This problem is exactly solvable at $\beta=2$ for a variety of background
potentials.  We introduce a new method of solving the problem, based on
mapping to a non-hermitian field theory, which may be easier to deal with
for certain potentials.  In addition, we look at a perturbation theory which
permits a mean-field analysis of the phase transition in this theory.

\section{Fermion Mapping at $\beta=2$}
As a a statistical mechanics system, the model, for a finite
number of particles in a uniform background, is defined by the partition 
function
\be
\label{Z}
\int (\prod \limits_{i=1}^{N} dz_i \, d\overline {z_i}) \prod \limits_{i<j}^{N} 
|z_i-z_j|^{\beta} \prod \limits_{i=1}^{N} e^{-\rho |z|^2}
\ee

We note that the factor $ \prod \limits_{i<j}^{N} |z_i-z_j|^{\beta}$ 
in the partition function is equivalent to a
correlation function in a free bosonic field theory:
\be
\langle \prod \limits_{i=1}^{N} e^{i\sqrt{4\pi} \Phi(z_i)} 
e^{-i N \sqrt{4\pi} \Phi(\infty)} \rangle
\ee
where the field $\Phi$ has the action 
$S=\frac{1}{\beta} \int dx dt (\nabla \Phi)^2$.

We may then go to an infinite number of particles, and write the partition
function as
\be
\label{zft}
Z=\int [d\Phi] e^{-S +\int e^{i\sqrt{4\pi} \Phi(x,t)} -i\rho \sqrt{4\pi}
\Phi(x,t) \, dx \, dt}
\ee
It may be seen that, by perturbatively expanding the partition function in
powers of $e^{i\sqrt{4\pi} \Phi}$, we recover 
the original statistical mechanics problem.

This equivalence between eqs. (\ref{Z}) and (\ref{zft}) is similar to the
equivalence between the statistical mechanics of an plasma consisting of
both plus and minus charges and the field theory of the sine-Gordon equation.

As is well known, the sine-Gordon equation at $\beta=2$ maps onto a problem
of free massive fermions, while at other temperatures the equation maps
onto the Thirring model, a model of interacting massive fermions\cite{coleman}.
We will
follow an analogous procedure in this case, leading to a non-hermitian
field theory of fermions.  All correlation functions will be given at
$\beta=2$, while a perturbation theory in a four-fermion interaction will
lead to other temperatures.

Each term of the bosonic action translates into a given term of an action
for relativistic fermions.  The action $S$ for the field $\Phi$
translates into $2\int dx \, dt \psi^{\dagger}_R \partial_{\overline z}
\psi_R +  \psi^{\dagger}_L \partial_{z} \psi_L$.  The term 
$ e^{i\sqrt{4\pi} \Phi(x,t)} $
translates into $2\pi a \psi^{\dagger}_L \psi_R$, where $a$ is an ultraviolet
cutoff for the theory.  
The existence of only one kind of charge in the
statistical mechanics theory leads to a non-hermitian field theory.

It is seen that placing a charge in the statistical mechanics theory at a
given point corresponds to turning a fermion at that point from a right-mover
into a left-mover.  The effect of the neutralizing background charge is to
turn left-movers back into right-movers via the anomaly.  One may imagine the
neutralizing background charge as a charge at infinity.  We can write
$i\Phi$ as $i \Phi \partial_t t$ and integrate by parts
to turn this into $i t \partial_t \Phi$ plus a boundary term.  This then
bosonizes into $\sqrt{\pi} t J_x$, which is $\sqrt{\pi}t (\psi^{\dagger}_R 
\psi_R- \psi^{\dagger}_R \psi_R)$.  
This has the effect of introducing a time-dependent
chemical potential which has different signs for the right- and left-movers.
If we imagine following the Hamiltonian evolution of the fermionic field theory
in imaginary time, right-moving particles are constantly converted into
left moving particles, but the energy of the states keeps changing so that
on average the number of holes in the negative energy sea is the same for the
both chiralities.  The effect of the boundary term resulting from the 
integration by parts is that we must start at $t=-\infty$ with a state in
which all right-moving states (of both positive and negative energy)
are filled and all left-moving states are empty and end at $t=\infty$ with
a state in which all left-moving states are filled and all right-moving
states are empty.  Let us refer to the state at $t=-\infty$ as the
state $V^-$.  We will refer to the state at $t=+\infty$ as the state $V^+$.

The final action for the fermion field is
\be
\int dx \, dt \, 2 (\psi^{\dagger}_R \partial_{\overline z}
\psi_R +  \psi^{\dagger}_L \partial_{z} \psi_L)
+\sqrt{2 \rho}
\psi^{\dagger}_L \psi_R
+\rho t 2 \pi ({\psi^{\dagger}_L \psi_L-
\psi^{\dagger}_R \psi_R})
\ee
There are several ways of writing the background charge which correspond to
choosing different gauges.
The factor $\sqrt{2 \rho}$ is chosen to cancel a factor arising later
from an integral of $\int dt e^{-2 \pi \rho t^2}$.  It also serves to give the
term in the action the correct dimensions.  This factor is unimportant since the
fixed density of the background charge means, due to charge neutrality, that
the number of particles in the statistical mechanics problem is fixed and thus
this factor just multiplies the partition function but does not affect
the physics. 
\section{Correlation Functions at $\beta=2$}
We compute correlation functions for the fermionic field and use them to
then obtain correlation functions in the statistical mechanics problem.  It
will hopefully be clear in which theory a correlation function is being
calculated.

We will compute propagators for the fermion field as follows: we 
will first compute the propagators for the case in which the 
two operators lie on the line $t=0$.  Then we will, in an appropriate gauge
for the background charge, generalize to arbitrary position of the
two operators.

In the operator formalism, we can write a two-point correlation function of
fields $\psi(1),\psi(2)$ as the expectation value
\be
\label{gf1}
\frac{1}{Z} \langle V^+|e^{-\int \limits_{0}^{\infty}H(t) dt} \psi(1) \psi(2) 
 e^{-\int \limits_{-\infty}^{0}H(t) dt} |V^- \rangle
\ee
where 
\be
Z= \langle V^+|e^{-\int \limits_{0}^{\infty}H(t) dt}
 e^{-\int \limits_{-\infty}^{0}H(t) dt} |V^- \rangle
\ee

The term in the Hamiltonian, $\int dx\, dt \psi^{\dagger}_L \psi_R
$, can be written in Fourier components as
$\int \frac{dk}{2\pi} 
a^{\dagger}(k)_L a(-k)_R$.  Each term in the integral is an eigenoperator
of the rest of the Hamiltonian, with eigenvalue $2k +4\pi\rho t$,
and so it is easy to rewrite eq. (\ref{gf1})
in terms of operators acting just at time $t=0$ as follows
\be
\label{gf}
\frac{1}{Z}\langle V^+|
e^{-\int \frac{dk}{2\pi} (\int\limits_{0}^{\infty} e^{-2 \pi \rho t^2 +2kt} dt )
\sqrt{2 \rho} a^{\dagger}(k)_L a(-k)_R}
 \psi(1) \psi(2)
e^{-\int \frac{dk}{2\pi} (\int\limits_{-\infty}^{0} e^{-2 \pi \rho t^2 +2kt} dt )
\sqrt{2\rho} a^{\dagger}(k)_L a(-k)_R}
|V^- \rangle
\ee

The simplest two-point function to compute is $\langle a^{\dagger}(k)_L a(-k)_R
\rangle$.  In the expectation value of eq. (\ref{gf}), for each $k$ the
operators $a^{\dagger}(k)_L$ and $a(k)_R$ must appear exactly once.  
It is clear then that 
$ Z=\prod \limits_{k} e^{\int \limits_{-\infty}^{\infty}
\sqrt{2\rho} e^{-2 \pi \rho t^2 +2kt} dt}$.  This is
\be
\label{pro}
\prod \limits_{k} e^{\frac{k^2}{2 \pi \rho}}
\ee
By inserting the operator $a^{\dagger}(k)_L a(-k)_R $ into the expectation 
value, we remove one term from the product of eq. (\ref{pro}).  This term is
$e^{\frac{k^2}{2 \pi \rho}}$.  Therefore, $\langle a^{\dagger}(k)_L a(-k)_R
\rangle =e^{-\frac{k^2}{2 \pi \rho}}$.  In real space this is $
\sqrt{\frac{\rho}{2}}e^{-\frac{\pi}{2} \rho x^2}$, 
and in an appropriate gauge it becomes $\sqrt{\frac{\rho}{2}}
e^{-\frac{\pi}{2}\rho |z|^2}$.

We next compute the function $\langle a^{\dagger}(k)_L a(-k)_L$.  
We note that initially all left-moving states are unoccupied at $t=-\infty$.
Then all states are occupied at $t=\infty$.  If the state with momentum $k$
is occupied at
$t=0$ then this gives a contribution of $1$ to the desired two-point function.
If it is unoccupied, then there is a contribution of $0$.  Therefore the
desired two-point function is
\be
\frac {\int \limits_{-\infty}^{0} e^{-2 \pi \rho t^2 +2kt} dt}
{\int \limits_{-\infty}^{\infty} e^{-2 \pi \rho t^2 +2kt} dt}
\ee
In real space, this is $\frac{1}{2 \pi i z} e^{-\pi \frac{\rho}{2}
|z|^2}$, where an appropriate gauge must be used to extend correlation 
functions off the line $t=0$.

The function in which both left-moving operators are replaced by right-moving
operators is similar.  The result is $\frac{1}{2 \pi i \overline z} 
e^{-\pi \frac{\rho}{2} |z|^2}$.

The most interesting two-point function to compute is
$\langle a^{\dagger}(k)_R a(-k)_L\rangle$.  This requires that one 
$a^{\dagger}(k)_L a(-k)_R$ operator appear at time $t<0$ and one must appear
at time $t>0$.  Therefore, the two-point function is
\be
\sqrt{2\rho}
\frac {
\int \limits_{-\infty}^{0} e^{-2\pi \rho t^2 +2kt} dt'
\int \limits_{0}^{\infty} e^{-2\pi \rho t'^2 +2kt'} dt}
{\int \limits_{-\infty}^{\infty} e^{-2\pi \rho t^2 +2kt} dt}
\ee

In real space, this is $
2 \rho \int \limits_{-\infty}^{0} e^{-2\pi \rho t'^2} dt'
\int \limits_{0}^{\infty} e^{-2\pi \rho t^2} dt
\int \frac{dk}{2\pi} e^{-\frac{k^2}{2 \pi \rho}}
e^{k(ix+2t+2t')}$.   
That becomes 
\be
\rho \int \limits_{-\infty}^{0} e^{-2 \pi\rho t'^2} dt'
\int \limits_{0}^{\infty} e^{-2 \pi \rho t^2} dt \,
e^{\frac{\pi \rho}{2}(ix+2t+2t')^2}
\sqrt{2 \rho}
\ee
This is equal to $\sqrt{2}\rho^{\frac{3}{2}}
\int \limits_{-\infty}^{0} dt'
\int \limits_{0}^{\infty} dt
e^{\frac{\pi \rho}{2}(-x^2 +8tt' +4ix(t+t'))}
$.
Integrate $t'$ to obtain
$\frac{1}{2\pi}\sqrt{\frac{\rho}{2}} \int \limits_{0}^{\infty} dt 
\frac{1}{t+\frac{ix}{2}}
e^{\frac{\pi \rho}{2}(-x^2 +ix4t)}$.
This is equal to
$\frac{1}{2\pi}\sqrt{\frac{\rho}{2}}  e^{+\frac{\pi \rho}{2}x^2}
\int_{\pi \rho x^2}^{\infty} e^{-t}/t$
which becomes
\be
\frac{1}{2\pi}\sqrt{\frac{\rho}{2}} e^{+\frac{\pi \rho}{2}|z|^2}
\rm{Ei}(\pi \rho |z|^2)
\ee
where $\rm{Ei}$ is the exponential integral function.

Having computed the propagators, it is trivial to compute correlation functions
in the statistical mechanics theory.
Let us compute the two-point particle correlation function at $\beta=2$.  
This is obtained by the expectation value $\langle 
\sqrt{2 \rho} \psi^{\dagger}_L(0) \psi_R(0) \sqrt{2 \rho}
\psi^{\dagger}_L(z) \psi_R(z)
\rangle$.
There are two ways to pair off the operators in this expression using
Wick's theorem.  The result is
\be
\rho^2 (1-e^{-\pi \rho |z|^2})
\ee
which agrees with the known result.

\section{Perturbation Theory}
By adding to the action of the fermionic theory a term of the form
$(\overline \psi \psi)^2$, we will, due to the bosonization rules, change
the value of $\beta$.  An attractive interaction will raise $\beta$, while
a repulsive interaction will lower $\beta$.  We will employ a simple mean-field
type procedure: the interaction will be decoupled into an interaction with
an external gauge field.  Then, we will integrate out the fermionic field
to obtain an action for the gauge field.  It will be found that at sufficiently
high, finite $\beta$ the quadratic term in the action for the gauge field will
change sign.  This will be identified as the location of a second order
phase transition.  It will then be argued that cubic terms appear in the
action for the gauge field, and will convert the phase transition to first
order.

For a given value of $\beta$, the desired fermionic action is
\be
\int dx \, dt \, 2 (\psi^{\dagger}_R \partial_{\overline z}
\psi_R +  \psi^{\dagger}_L \partial_{z} \psi_L)
+\sqrt{2 \rho}
\psi^{\dagger}_L \psi_R
+\rho t 2 \pi ({\psi^{\dagger}_L \psi_L-
\psi^{\dagger}_R \psi_R})
+ (\frac{1}{\beta}-\frac{1}{2}) 4 \pi 
:\psi^{\dagger}_R \psi_R:  :\psi^{\dagger}_L \psi_L:
\ee
We can write the interaction term by introducing an external gauge field
$A_{R,L}$.  The action for the gauge field plus the coupling between the
gauge field and the fermion field is
\be
\int dx\, dt\, A_R J_R + A_L J_L + \frac{1}{4\pi}
\frac{1}{\frac{1}{2}-\frac{1}{\beta}} A_R A_L
\ee
where we define $J_R = :\psi^{\dagger}_R \psi_R: $ and
$J_L = :\psi^{\dagger}_L \psi_L: $ and where $A_{R}$ is the complex conjugate
of $A_L$.

It will turn out, when we compute various diagrams, that results will not
be gauge invariant.  This is unrelated with the particle choice of gauge
we made to compute propagators, and instead is a result of the boundary
conditions due to the neutralizing background field.  The theory is also
not invariant under charge conjugation.  This implies that Furry's 
theorem\cite{furry} need not hold, and we may obtain contributions from
fermion diagrams with an odd number of photon vertices attached.  This will
then convert the second order transition into a weakly first order transition.

Let us first compute the quadratic terms in the action for $A_{R,L}$ resulting
from integrating out $\psi$.  There are three different types of diagrams which
must be considered, which involve computing different 
current-current correlation functions.  We may have to compute a correlation
function involving two currents of the same chirality.  This will be considered
first.  We may have to compute a correlation function involving currents of
the opposite chirality.  This correlation function splits into two
pieces: one piece which may be obtained by naively calculating diagrams, and
one piece which is a result of the need for a regulator field to get rid
of divergences.

The introduction of the quadratic term will change the action of the
gauge field, as a function of wavevector $k$, to
\be
\begin{array}{l}
\frac{1}{4\pi}\frac{1}{\frac{1}{2}-\frac{1}{\beta}} A_R(k) A_L(-k)
-\frac{1}{2}\langle J_R(k) J_R(-k) \rangle A_R(k) A_R(-k)\\
-\frac{1}{2}\langle J_L(k) J_L(-k) \rangle A_L(k) A_L(-k)
-\langle J_R(k) J_L(-k) \rangle A_R(k) A_L(-k)) \end{array}
\ee
All that needs to be computed now are some current-current correlators.

The current-current correlation function $\langle J_R(0) J_R(z) \rangle$
is equal to $\frac{1}{(2\pi z)^2}e^{-\pi\rho|z|^2}$.  
We must take the Fourier transform
of this at a given momentum $k$.  In order to do this, it is necessary
to introduce a massive regulator field.  Without the term $e^{-\pi\rho|z|^2}$
in the correlator, the Fourier transform of $1/z^2$ is known to be
$\frac{1}{4\pi}\frac{k_L^2}{|k|^2}$, where $k_L$ is equal to $k_x-ik_t$,
$k_R$ is equal to $k_x+i k_t$, and
$|k|^2=k_L k_R$.  The term
$e^{-\pi\rho|z|^2}$ multiplies the rest of the correlator in real space and
thus convolves with the correlator in Fourier space.  The $k$-th
Fourier component of the current-current
correlator in momentum space is then 
$\int dk_L dk_R \frac{1}{4\pi}\frac{l_L}{l_R}
e^{-\frac{|k-l|^2}{4\pi\rho}}$.  This integral can be performed analytically and
the result is $\frac{1}{4\pi}\frac{k_L^2}{|k|^2} 
(1-e^{-\frac{|k|^2}{4\pi\rho}})$.
It is seen that at large $k$ the current-current correlator is unchanged from
the correlator for a fermionic system with no non-hermitian term.
The correlation function of two left-moving currents is calculated in the
same way, simply replacing $k_L$ by $k_R$ and vice-versa.

The naive contribution to the correlator of two currents of opposite
chirality is given by $\langle J_R(0) J_L(z)\rangle$.  This is
$\frac{\rho}{4\pi}\rm{Ei}(\pi \rho |z|^2)$.  The Fourier transform is
$\frac{\rho}{4\pi} \int dx \, dt \, \int \limits_{1}^{\infty} \frac{da}{a}
e^{ikx-\pi \rho a(x^2+t^2)}$.  Doing the integral over $x$ and $t$ first,
we obtain $\frac{1}{4\pi} \int \limits_{1}^{\infty} \frac{da}{a^2}
e^{-\frac{|k|^2}{4 a \pi \rho}}$.  Changing from $a$ to $1/a$, this equals
$\frac{\rho}{k^2} (1-e^{-\frac{k^2}{4 \pi \rho}})$.

The necessity of introducing a massive regulator field to compute the
correlation function of two currents of the same chirality gives an
additional contribution to the correlation function of two currents of 
opposite chirality.  This piece is infinitely short range, and so independent
of $k$.  It is equal to $\frac{1}{4\pi}$.  
For the fermionic system with no non-hermitian term, such a piece
is needed to maintain gauge invariance.

Even without looking closely at the Fourier transform
of the current-current correlators, we may easily see that there will be
an instability for finite $\beta$ at the level of the quadratic action for
$A_{L,R}$.  For the fermionic system with no non-hermitian term, the
quadratic term changes sign, for {\it all} 
$k$ transverse to the direction of the
gauge field, at $\beta=\infty$, which corresponds to a finite value of the
attractive interaction.  By adding the non-hermitian terms to the action,
we have altered the quadratic terms.  The correlation function of two
currents of the same chirality has been reduced, but the reduction decays
exponentially for large $|k|^2$.  The correlation function of two currents
of the opposite chirality has been increased by an amount which decays
only algebraically for large $|k|^2$.  Thus, for sufficiently large
$|k|^2$, the term which is added to the action of the $A_{L,R}$ field is
increased and the intability will occur at a lower value of the
attractive interaction.  This corresponds to an instability at a
finite value of $\beta$.

Let us locate the transition temperature in this lowest order theory.
The response to a gauge field is strongest when the field is transverse.
The response then is (summing all different terms and taking $A_R=A_L$)
\be
\frac{1}{4\pi} + \frac{1}{4\pi}(1-e^{-\frac{k^2}{4\pi\rho}})
+\frac{1}{k^2}(1-e^{-\frac{k^2}{4\pi\rho}})
\ee
This function is equal to $\frac{1}{2\pi}$ at $k=0$ and $k=\infty$.  The
function increases in magnitude as $k$ increases from $0$, until it hits
a maximum, then decreases again.  Numerically, the maximum is at 
$k=4.74711 \sqrt{\rho}$.  At this wavevector, we find that the theory goes
unstable at a temperature of $\beta=15.4036$.

In a mean-field approximation, in the absence of cubic terms, the instability
discussed in the quadratic action for $A_{R,L}$, would lead to a second order
transition.  It will now be shown that cubic terms are non-vanishing and
that this transition becomes first order.  
Due to the complexity of the diagrams, we will not
explicitly compute the cubic and quartic terms, and will simply show that the
cubic term is non-vanishing.

In 2 dimensions Furry's theorem is very easy to prove for a massive or
massless Dirac field.  If we have a fermion
loop with an odd number of vertices, we may imagine reversing the direction
of the propagation of the fermion around the loop.  This will change the sign
of all propagators which preserve chirality (propagators of the form
$\langle \psi^{\dagger}_R \psi_R \rangle$ or $\langle \psi^{\dagger}_L \psi_L
\rangle$) of the fermion field, and 
preserve the sign of all propagators that change chirality.  Since the fermion
must have the same chirality after it goes around the loop, there are an even
number of propagators which change chirality, and thus an odd number of 
propagators which preserve chirality.  This means that the total sign of the
diagram changes when reversing the direction of propagation around the loop and
thus the total contribution from both directions is zero.

For the non-hermitian theory considered here, the above proof breaks down.
When reversing the direction of propagation around the loop, we may change
some propagators from $\langle \psi^{\dagger}_L(z_1) \psi_R(z_2) \rangle$ to
$\langle \psi^{\dagger}_R(z_2) \psi_L(z_1) \rangle$.  
Since the non-hermitian theory
lacks charge conjugation invariance, these two propagators have different
values, and thus the two contributions do not cancel.  This means that
cubic terms do appear in the action for $A_{R,L}$ and the transition becomes
first order.  It should be noted that these cubic terms are intrinsically
not gauge invariant, since they vanish in a gauge in which $A_L=0$ or $A_R=0$.
\section{Extension to Curved Space}
We will also consider this model on a space of constant negative curvature.
We may define such a space in the half-plane given by $Im(z)>0$, or
equivalently, $t>0$.  We will
use a conformal gauge for the metric such that $ds^2=\frac{2}{t^2}
(dx^2+dt^2)$.  Then the scalar curvature $R=-1$.

It is easy to transcribe the action for the bosonic field to this curved
space and the result is
\be
Z=\int [d\Phi] e^{-S +\frac{1}{t^2}
\int e^{i\sqrt{4\pi} \Phi(x,t)} -\frac{2i}{t^2}\rho \sqrt{4\pi}
\Phi(x,t) \, dx \, dt}
\ee

When going to a fermionic action, one must be slightly careful.  The curved
space means that in order to maintain the same ultraviolet cutoff $a$ when
measured in length $\sqrt{ds^2}$
everywhere, the ultraviolet cutoff when measured in length $\sqrt{dx^2+dt^2}$
must vary proportional to $t$.  Therefore, the operator 
$\int e^{i\sqrt{4\pi} \Phi(x,t)}$ bosonizes into something proportional to
$(2\pi t) \psi^{\dagger}_L \psi_R$

This means that the final fermionic action for the curved space problem is
\be
\int dx \, dt \, 2 ({\psi^{\dagger}_R \partial_{\overline z}
\psi_R +  \psi^{\dagger}_L \partial_{z} \psi_L})
+\frac{1}{t}
\psi^{\dagger}_L \psi_R
+4 \pi \frac{\rho}{t}  (\psi^{\dagger}_R \psi_R-
\psi^{\dagger}_L \psi_L)
\ee
In this action, the density, $\rho$, is dimensionless, as we have
picked units in which the curvature is one.  We can use $1/\rho$ to
measure the curvature of the manifold in units of particle spacing squared.
As $\rho \rightarrow
\infty$, we expect to recover the flat space results.

It is now possible to proceed as before and calculate propagators.  We will
simplify and sketch the calculation of only $\langle \psi^{\dagger}_L \psi_R
\rangle$.  We restrict to the line $t=1$.  The only new feature that
emerges is that the final state is different.  The initial
state, $|V^- \rangle$, is still a state in which all right-moving states
are occupied and all left-moving states are empty.  The system starts in
this state at time $t=0$.  However, there is no term left from the
integration by parts at $t=\infty$ and therefore the system ends in the
vacuum state $|V\rangle$ which is the normal vacuum state for a
the free fermion theory, where all negative energy states are filled and
all positive energy states are empty.

Following a similar procedure as before, we move all 
$\psi^{\dagger}_L \psi_R$ operators resulting from the Hamiltonian evolution 
to the time $t=1$.
There, an operator $a^{\dagger}(k)_L a(-k)_R$ comes with the amplitude
\be
\int \limits_0^{\infty} dt \frac{1}{t} 
e^{-\int \limits_{t}^{1} 2k +8\pi\rho/t' dt'}
\ee
This is equal to
$\int \limits_{0}^{\infty} dt \frac{1}{t^{1-8\pi\rho}}
e^{2k(t-1)}$, which is proportional to $e^{-2k}\frac{1}{|k|^{8\pi\rho}}$.  
Therefore,
$\langle a^{\dagger}(k)_L a(-k)_R \rangle$ at $t=0$ is proportional to
$e^{2k} |k|^{8\pi\rho}$.  
Due to the initial and final states, $|V^- \rangle$ and
$|V \rangle$, the propagator is non-vanishing
only for $k<0$, when
this Fourier transforms to $\frac{1}{(1-i\frac{x}{2})^{1+8 \pi\rho}}$.

The result for the correlation function in the statistical mechanics problem, 
obtained from calculating
$\langle \psi^{\dagger}(0)_L 
\psi(0)_R 
\psi(x)_L 
\psi(x)_R 
\rangle$, is
\be
1-(\frac{1}{1+\frac{x^2}{4}})^{1+8 \pi \rho}
\ee
If we take a limit as $\rho$ goes to infinity, and appropriately rescale
$x$ to measure length in the local particle spacing, the correlation function
turns back into a Gaussian, as in the flat space case.
Recall that the distance between two points on the line $t=1$, located
at $x=0$ and $x=x$ is, measured in the metric for the negative curvature
space, less than $x$.  In reality it goes logarithmically with $x$ for
large $x$.  Thus, the actual decay of correlations in the curved
space problem is exponential, as would be expected from a high-temperature
expoansion.
The slower decay of the correlations in the curved space case, exponential
instead of Gaussian, indicates that screening is less effective than in
flat space.
\section{Conclusion}
A new formulation has been given of the one-component plasma problem.  This
provides an alternative way of deriving old results.  This technique may
turn out to be simpler than others for certain background potentials.
In addition, the perturbation theory of this model in terms of fermionic
operators is very different from the previously developed perturbation
theory for the statistical mechanics theory\cite{pert}, and may lead to
easier calculations.

In particular, we located an instability of the theory at the quadratic level.
It must be noted that the location of the second order transition in this
theory, at $\beta=15.4036$, is approximately an order of magnitude lower
than the actual location of a first order transition.  Also, the
appearance of cubic terms, while converting the second order transition
to first order, will further lower the transition $\beta$.  This is
evidence that the higher order corrections to the effective action for
the gauge field are not negligible.  However, the manner in which the
transition temperature is calculated makes this number very sensitive to
small adjustments in the effective action; one must calculate the response
functions in the fermionic theory and then relate $\beta$ to the
reciprocal of the change in the action for the gauge theory.  The act
of taking the reciprocal makes this procedure less accurate.  Unfortunately,
we are also unable to understand the particular wavevector which goes
unstable, as we do not see how to easily relate this wavevector to any
spacing in a triangular lattice.  In addition, since the order parameter is
a vector, it is possible to construct a mean-field state that breaks 
orientational symmetry but not translational symmetry
(constant non-vanishing gauge field).  This would require that the
first wavevector to go unstable would be at $k=0$.  This does not happen
yet at the one-loop level, though one would expect it would happen to higher
orders.

However, we may hope that the perturbation series for the effective action
of the gauge field will be convergent, order by order in $(1/2-1/\beta)$, as
the only actual instability of the theory is at $\beta=\infty$.  This implies
a radius of convergence of $1/2$, and thus the theory should converge for
$1<\beta<\infty$.  Arguments like Dyson's instability argument for the 
non-convergence of the perturbation series do not apply here since there is no 
instability near $\beta=2$\cite{dyson}.  To make this convergence more
clear, it may help to adjust units so that the action for the gauge field is
\be
\int dx\, dt\, A_R A_L + 
\sqrt{4\pi (\frac{1}{2}-\frac{1}{\beta})} (A_R J_R + A_L J_L)
\ee
to avoid what may look like a singularity at $\beta=2$.
Thus, a sufficiently high order calculation in
this theory should yield the effective action for the gauge field, which
can then be treated in a mean-field fashion to, in principle, extract
the transition properties with arbitrary accuracy.  Of course, the series
might diverge at the point of a second order phase transition due
to infrared problems, but since a
first order phase transition is expected to occur before the second order
transition, this is not a problem.

Finally, the model has been considered on a curved manifold.  The two-point
correlations in the original statistical model have an exponential decay instead
of a Gaussian decay.  Since the decay of this quantity reflects the effects
of screening, it seems that on a curved manifold the system screens less well.
This may be due to an effect of frustration, introduced by the curvature.
It would be interesting to extend the perturbation theory to a curved manifold,
both to look at the RPA as well as to look at correlation functions away
from $\beta=2$.  It is believed from a perturbation theory for the original
statistical mechanics model\cite{pert} that the correlation functions on
a flat space begin to show short-range order as soon as $\beta>2$.  This
means a lowest order perturbation theory calculation for the curved space
may show interesting frustration effects.

\end{document}